\documentclass[reprint,fleqn,superscriptaddress,twocolumn,showpacs,
amsmath,amssymb,prb,aps,bibliography]{revtex4-1}
\usepackage[all]{xy}
\usepackage{gensymb}
\usepackage{graphicx,psfrag,times,epsfig,color}
\usepackage{verbatim,natbib}
\usepackage{color}
\usepackage{makeidx}
\usepackage{amsmath}
\usepackage{bm}
\usepackage{amsfonts}
\usepackage{amssymb}
\usepackage{hyperref}

\begin{document}

\title{Evidence that cuprate superconductors form an array of nanoscopic Josephson junctions}
\author{H\'ercules S. Santana and E. V. L. de Mello}
\affiliation{Instituto de F\'{\i}sica, Universidade Federal Fluminense, 24210-346 Niter\'oi, RJ, Brazil}

\email[Corresponding author: ]{evandro@mail.if.uff.br}
\email[Corresponding author: ]{hercules\_santana@id.uff.br}

\begin{abstract}

Recent measurements of charge instabilities in overdoped compounds rekindled the proposal that cuprates
become superconductors by long-range order through Josephson coupling between nanoscopic charge domains.
We use the theory of phase-ordering dynamics to show that incommensurate charge density waves (CDWs) are
formed in the CuO planes by a series of free-energy wells separated by steep barriers. Charge oscillations in
these domains give rise to a net hole-hole attraction proportional to the height of these barriers. Concomitantly,
the self-consistent calculations yield localized superconducting amplitudes in the CDWdomains characterizing a
granular superconductor. We show that a transition by long-range phase order promoted by Josephson coupling
elucidates many well-known features of cuprates like the high magnetic penetration depth anisotropy and the
origin of the pseudogap, among others. Furthermore, the average Josephson energy reproduces closely the planar
superfluid density temperature dependence of La-based films and the superconducting giant proximity effects of
cuprates, a 20-year-old open problem.
\end{abstract}
\maketitle
\section{Introduction} 

One of the main challenges of condensed-matter physics is a complete theory for high critical temperature superconductors (HTSs). Such a theory has been hindered by many issues like the absence of clear Fermi surfaces, the pseudogap, and the prominence of various forms of collective fluctuations\cite{Keimer2015}. 
To characterize whether these distinct orders compete with or strengthen each other, many experimen tal techniques have been refined to detect even overlapping fluctuations. After imaging a granular structure with high spatial resolution scanning tunneling microscopy (STM) in underdoped Bi$_2$Sr$_2$CaCu$_2$O$_{8+d}$ (Bi2212), 
Lang {\it et al}\cite{Lang2002} proposed that superconducting (SC) long-range order could be achieved 
by Josephson coupling between nanoscopic domains.
This idea gained more recognition after the measurements of CDW in 
YBa$_2$Cu$_3$O$_{6+x}$ (YBCO) single crystals but was still not 
considered as a general theory
of cuprates for two reasons: first, the absence of 
incommensurate charge ordering (CO) or CDW data in the overdoped region and, second,
the lack of a theoretical model that could 
justify and describe the physical formation of the small Josephson junctions between
the CO domains.

In this paper, we use the theory of phase-ordering dynamics to show that the
CDW or CO may be formed by a two-dimensional array of free energy potential wells with
similar properties of granular superconductors.
In recent years there was a great improvement in the precision of 
the CO wavelength $\lambda_{\rm CO}$ measurements by STM, x-ray, 
and Resonant X-Ray Scattering (REXS)\cite{Comin2016}.
The very fine variation of $\lambda_{\rm CO}$ with the doping $p$ (or
hole per CuO unit cell) revealed in these
experiments can be reproduced theoretically 
by a phase separation formalism based on the time-dependent nonlinear Cahn-Hilliard (CH)
differential equation\cite{Cahn1958}. In this approach, the charge modulations maybe tuned 
up to reproduce the measured $\lambda_{\rm CO}(p)$ on 100\% volume
fraction of the simulations\cite{deMello2009,DeMello2012,deMelloKasal2012,DeMello2014,Mello2017}.

The phase separation free energy
reproduces the CDW structure and acts as a crystal field
that promotes the Cooper pair formation,
leading to a direct connection between the SC interaction and the charge modulations. 
Some critique the CDW-mediated superconductivity, arguing that CDW is limited
to the underdoped region.  However
recent measurements\cite{Wu2017,Shen2019,OverFei2019} and new 
x-ray diffraction demonstrated CDW correlations in overdoped 
La$_{2-x}$Sr$_x$CuO$_4$ (LSCO) up to compounds of at least $x \equiv p = 0.21$
and possibly up to $p = 0.25$\cite{Tranquada2021}. 
Here we perform CDW simulations and develop a
SC theory of cuprates based on Josephson coupling between local SC order parameters 
in these domains and their long-range phase order (LRO). 
The Josephson coupling is closely related with the superfluid density\cite{Spivak1991,Mello2021} 
$\rho_{\rm sf}$ and reproduces its measured temperature variation $\rho_{\rm sf}(T)$
of several overdoped LSCO films with great accuracy.

The CDW-LRO approach is also appropriate to describe the giant proximity effects
(GPE) experiments in YBCO S-I-S wires\cite{ybco2000PE} and in
LSCO Josephson type trilayers S-N'-S junctions\cite{Bozovic2004,Morenzoni2011} 
where I is an insulator with nonzero doping, S is a superconductor,
and N' is a superconductor layer in the normal phase.
According to conventional theory, the critical current should diminish exponentially
with the size $d$ of a barrier made of non-SC materials\cite{Bozovic2004}. 
For traditional low temperature Josephson junctions
the thickness of the barrier $d$ is generally
comparable with the barrier coherence length\cite{Bozovic2004} $\xi_{\rm N}$. 
However, despite
the very small  SC coherence lengths $\xi_{\rm SC}$ 
of cuprates, GPE was measured with
several large barriers\cite{ybco2000PE,Bozovic2004} with
$d \gg \xi_{\rm N}$. Thus, technically the YBCO non-SC
spacer with $d \sim 100$ nm and La-based trilayers with $d \sim 100-1000$ nm are too
large to carry a critical current, in clear contradiction 
with the experiments\cite{ybco2000PE,Bozovic2004,Morenzoni2011}. On the other hand,
in our approach, S, N', and I all have CDW and differ only by the presence or absence of 
LRO, which is very sensitive to external perturbations like
an applied current or magnetic field.

\section{The CDW simulations}

The starting point is the definition of the time-dependent
phase separation (PS) order parameter associated with the local electronic
density, $u({\bf r},t) = [p({\bf r},t) - p]/p$, where $p$ is the average
hole density and $p({\bf r},t)$ is the charge density at a position ${\bf r}$
in the CuO plane and at a time of simulation $t$.
The CH equation is based on the electronic phase separation
Ginzburg-Landau (GL) free energy expansion in terms of the conserved charge order parameter $u$    \cite{Otton2005,deMello2009,DeMello2012,deMelloKasal2012,DeMello2014,Mello2017}.
\begin{equation}
f(u)= {{\frac{1}{2}\varepsilon |\nabla u|^2 +V_{\rm GL}(u,T)}}\;,
\label{FE}
\end{equation}
where $\varepsilon$ is the parameter that controls the charge modulations scale
and ${V_{\rm GL}}(u,T)= -\alpha [T_{\rm PS}-T] u^2/2+B^2u^4/4+...$ is a 
temperature-dependent double-well potential that characterizes the rise of charge oscillations 
below the onset of phase separation temperature $T_{\rm PS}$. We do not know
$T_{\rm PS}$, but there are indications that it is close to the pseudogap temperature $T^*$.
In the  simulations, when  $T \le T^*$, the values of $\alpha$ and $B$ are always one. 
This free energy in terms of the phase separation order
parameter is much simpler than the Ginzburg-Landau-Wilson free energy in 
terms of SC and pair density wave fields (PDW)\cite{InterOrder2015},  but it suitably reproduces
the details of the CO structure of distinct compounds and their localization
energy ${V_{\rm GL}}$.

\begin{figure}[!ht]
 \includegraphics[width=8.0cm]{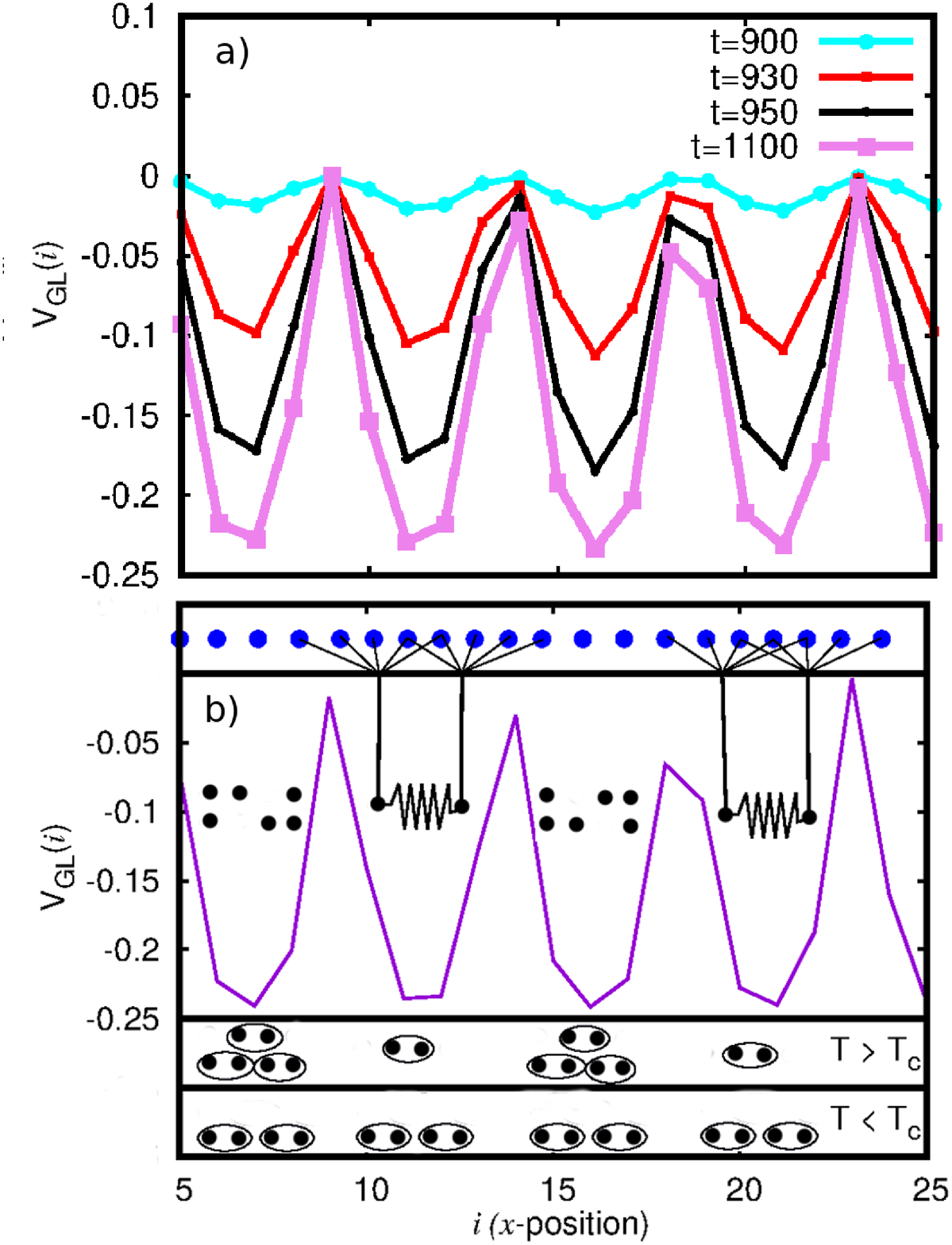}
\caption{ (a) The $V_{\rm GL}(t)$ evolution with time. 
Initially, $V_{\rm GL}$ is flat, which corresponds
to a system above the onset phase separation temperature $T_{\rm PS}$. 
The phase separation potential wells or valleys increase with time, and this 
behavior is correlated with the 
decrease of the temperature below $T_{\rm PS}$. As the temperature goes below 
$T^*$, we assume the stable large time behavior $V_{\rm GL}(t \rightarrow \infty)$ to
be that of low temperature $V_{\rm GL}(T \le T^*)$.  
(b) The Cooper pair formation in the CDW free energy valleys.
At the top, we represent some planar Cu atoms (blue-filled circles) attracted 
to hole-poor domains
represented by black lines as an illustration. Hole fluctuations in these domains produces 
atomic fluctuations that affect also the other holes, promoting an
atomic mediated interaction (represented by the springs as an illustration). 
At low temperature ($T \le T^*$) the Cooper pairs may be formed in the CDW valleys (the
encircled pair of black dots), and at $T \le T_c$ they superflow and become uniform
on the CuO plane (in agreement with the CO x-ray scattering decreasing signal below 
$T_c$\cite{Wise2008,Chang2012}).
}
\label{VGLtime.Temp}
\end{figure}

The CH equation can be written derived by a continuity equation 
of the local free energy current density  ${\bf J} = M{\bf\nabla}(\delta f/ \delta u)$,\cite{Bray1994}
\begin{eqnarray}
\frac{\partial u}{\partial t} & = & -{\bf \nabla.J} \nonumber \\
& = & -M\nabla^2[\varepsilon^2\nabla^2u
- \alpha^2(T)u+B^2u^3],
\label{CH}
\end{eqnarray}
where $M$ is the mobility or the charge transport coefficient that sets both the phase separation time scale and
the contrast between the values of $u$ for the two phases. 

The equation is solved by a stable and 
fast finite difference scheme  with free boundary
conditions\cite{Otton2005}, yielding the 
phase separation order parameter $u({\bf r},t = n\delta t)$,
function of position ${\bf r}$ and $n$ simulation time step $\delta t$.
The limiting cases are
$u({\bf r}_i,t)\approx 0$, corresponding to homogeneous systems above 
or near $T_{\rm PS}$
or small charge variations like the observed CDW, and
$u({\bf r}_i,t\rightarrow \infty) =  \pm  \:1$, corresponding to the extreme case 
(at low temperatures) of complete phase separation. 
The local charge density is derived from
$p({\bf r},t) = p \times (u({\bf r},t) + 1)$, and the latter case (strong phase separation) 
applies to static 
stripes\cite{Tranquada1995a,Thampy2017}, while the former (weak phase separation) to very small 
$\Delta p \approx 10^{-2-3}$ variations around $p$, like that measured 
in YBa$_2$Cu$_3$O$_{6+\delta}$ (Y123)\cite{Kharkov2016}.

Figure \ref{VGLtime.Temp}(a) shows that below  the $T_{\rm PS}$ temperature the 
order parameter $u({\bf r},t)$ evolves in time
and $V_{\rm GL}({\bf r}, t)$ valleys become deeper, which is expected to occur when the temperature
decreases, favoring the mesoscopic phase separation. As the temperature goes below 
$T^*$, we assume the large time behavior $V_{\rm GL}({\bf r}, t \rightarrow \infty)$
shown in Fig. \ref{VGLtime.Temp}(a) to become
the low temperature $V_{\rm GL}({\bf r}, T \rightarrow 0)$ that generates
the CDW [Fig. \ref{VGLtime.Temp}(b)]. In this large time regime, $V_{\rm GL}({\bf r})$ depends on
the temperature through the usual form of the first GL coefficient defined after Eq. \ref{FE},
that is, $(1 - T/T^*)^2$.
This temperature dependence is the only relevant dependence of $V_{\rm GL}({\bf r}, T)$ in our
calculations.

\begin{figure}[!ht]
 \includegraphics[height=8.0cm]{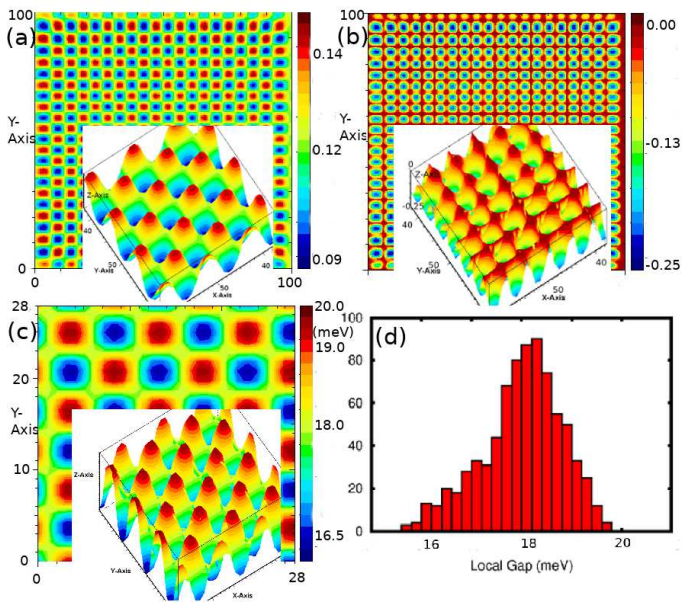}
\caption{ Top and three dimensional perspective of different properties at a later time or
low temperature:
(a) A checkerboard CDW map for a LSCO system with average hole density 
$p = 0.12$ on a $100\times100$ unit cell simulations and out of plane view. 
(b) The corresponding GL free energy potential $V_{\rm GL}({\bf r}_i)$. The inset shows
the three-dimensional perspective, which reveals the array of deep wells. 
(c) The BdG SC amplitude $\Delta_d({\bf r}_i)$ map in a $28\times28$ unit cell portion of the density map of (a). 
The three-dimensional perspective plot  demonstrates that the amplitudes $\Delta_d({\bf r}_i)$
are completely localized inside the CDW domains. 
(d) Typical histogram of the amplitudes shows the $\Delta_d({\bf r}_i)$ variation with
the local $p_i$ concentration.
}
\label{MapsHist3}
\end{figure}

A typical later time and low-temperature $V_{\rm GL}(u({\bf r}))$  used in the calculations
is shown in Fig. \ref{MapsHist3}(b), which leads to the LSCO checkerboard
structure. We also show the three-dimensional side view (in the inset)
with its wells or valleys in form of ``ice cream cones''. 
Notice that the free energy is defined over the CuO 
plane, and the third dimension is its values in each site,
which demonstrates that the minima occur at the center of the 
charge domains and the maxima at the borders. 
The detailed structure and strength of this potential is shown in the 
Supplemental Material\cite{Supplemental}.

For YBCO, the CDW domains are formed in puddles or patches with
different modulation in either {\it a} or {\it b}-direction in the CuO plane\cite{Comin2015a,YUnidire2021}.
The LRO superconducting calculations with $V_{\rm GL}(u({\bf r}))$ and alternating 
stripe-like puddles follow in the same way of the checkerboard of 
Fig. \ref{MapsHist3} since the method
requires only the formation of such finite-charge domains\cite{Mello2017}.
We emphasize that the calculations described and performed here 
are pure two dimensional in order to model the observed charge arrangements on the
CuO plane and does not take into account
possible out-of-plane influences, like the quenched disorder which arises 
from oxygen interstitials discovered in the in HgBa$_2$CuO$_{4+y}$ system\cite{Campi2015}.

\section{Superconducting Calculations}

The new form of analysis here is the three-dimensional view
of the planar CDW maps shown in Fig. \ref{MapsHist3}(a),
the GL free energy potential $V_{\rm GL}$ 
in Fig. \ref{MapsHist3}(b), and the SC pair amplitude $\Delta_d({\bf r}_i)$ in a $28\times28$ site 
portion of the $100\times100$ site density map from Fig \ref{MapsHist3}(a) in Fig. \ref{MapsHist3}(c). 
These illustrative plots are for an LSCO sample with an average hole density  $p = 0.12$. 
These plots show clearly that alternating rich and poor
charge regions develop in the same kind of valleys in the form of ``ice cream cones'', as 
it is demonstrated by the out-of-plane view of the insets.  The SC pair
amplitudes inside the wells in Fig. \ref{MapsHist3}(c) have the same properties
of isolated SC grains, which motivates the proposal of an array of Josephson junctions.

While Fig. \ref{VGLtime.Temp}(a) shows the evolution of $V_{\rm GL}(x_i)$ that leads 
to the CDW, Fig. \ref{VGLtime.Temp}(b) shows the low-temperature $V_{\rm GL}(x_i)$ 
that leads to the formation of the SC interaction in the CuO planes.
At the top we represent some planar atoms (blue circles) attracted (repelled)
by hole-poor (hole-rich) CDW domains
represented schematically by black lines.
High-energy x-ray diffraction\cite{Chang2012} revealed that CDW modulations displace 
the  Cu and O atoms whose oscillations around their
equilibrium positions are sensed by the holes and may give rise 
to a net hole-hole attraction\cite{Mello2017,Mello2021} illustrated by the springs. 

We argued before that this induced hole-hole SC interaction is proportional to  
the depth of the $V_{\rm GL}$ wells averaging over the whole system, that is, 
${\left < V_{\rm GL}(p)\right >}$\cite{Mello2017,Mello2021}.
The mean-field self-consistent Bogoliubov-de Gennes (BdG) calculations with this
attractive pair interaction over the whole system yield the local 
pair amplitude map $\Delta_d({\bf r}_i)$. 
The results also depend on the local hole density $p(i)$ and, consequently,
have the same CDW modulations ($\lambda_{\rm CO}$), 
leading to what is known as pair density wave \cite{Keimer2015}.

Since the $\Delta_d({\bf r}_i)$ fit 
perfectly inside the wells, each CDW domain has an independent local 
SC order parameter phase $\phi_i$, exactly as a granular superconductor. 
Therefore, each CDW domain may behave like mesoscopic SC grains
interconnected by Josephson junctions with  energies $E_{\rm J}({\bf r}_{lm})$ 
between $l$ and $m$ domains like those shown with alternate red and blue colors 
in Fig. \ref{MapsHist3}(c). In the bottom of Fig. \ref{VGLtime.Temp}(b) we show 
that, according to previous calculations\cite{Mello2017,Mello2021}, 
the Cooper pairs are present up to $T \le T^*$ and are represented by two
encircled black dots. At $T \le T_c$, LRO sets in, and the
Cooper pairs superflow through the system, establishing a uniform charge density
on the CuO plane. This leads to a decrease of the CDW signal observed by 
high-energy x-ray diffraction below $T_c$\cite{Wise2008,Chang2012} and was widely
interpreted by due to the competition between CDW and superconductivity. We
emphasize that our calculations establish the opposite: The CDW hosts the charge
domains and the local Cooper pairs in alternating domains, but they spread through 
the system when LRO sets in at $T \le T_c$. This is shown
schematically at the bottom of Fig. \ref{VGLtime.Temp}(b).

Following the above arguments, $T_{\rm c}$ is determined by the competition between
thermal disorder and the average planar Josephson energy
${\left < E_{\rm J}(p,T) \right >}$ that depends also on the
spatial averaged $d$-wave pair amplitude
\begin{equation}
{\left  <\Delta_d(p,T)\right >} = \sum_i^N \Delta_d({\bf r}_i,p, T)/N, 
\label{DdM}
\end{equation}
where
the sum, like in the case of ${\left < V_{\rm GL}(p)\right >}$, is over the N unit cells of a CuO plane.
As explained previously\cite{DeMello2014,Mello2017}, the $d$-wave relation for 
the average Josephson coupling energy is proportional to the 
$s$-wave expression\cite{AB1963}:

\begin{equation}
 {\left < E_{\rm J}(p,T) \right >} = \frac{\pi \hbar {\left <\Delta_d(p,T)\right >}}
 {4 e^2 R_{\rm n}(p)} 
 {\rm tanh} \bigl [\frac{\left <\Delta_d(p,T)\right >}{2k_{\rm B}T} \bigr ] .
\label{EJ} 
\end{equation}
Here $R_{\rm n}(p)$ is the average planar tunneling resistance between the grains
that is assumed to be proportional to the in-plane normal state resistance
just above $T_{\rm c}$. 
In the array of Josephson junction model, the current is composed of Cooper pairs
tunneling between the CDW domains
and by the normal carriers or quasi-particle plane current\cite{Bruder95}. 
For a $d$-wave HTS near $T_{\rm c}$ 
the supercurrent is dominant\cite{Bruder95}, which justifies the use of the
experimental $R_{\rm n}(p)$ in the above equation. 
Therefore, LRO is attained when the average ${\left < E_{\rm J}(p,T) \right >}$
is strong enough to overcome thermal phase disorder, or 
${\left < E_{\rm J}(p,T_{\rm c}) \right >} = k_{\rm B}T_{\rm c}$,
and that is how we
derive $T_{\rm c}$\cite{DeMello2014,Mello2017,Mello2020a,Mello2020c}.

These in-plane calculations are the basic pillars to the three-dimensional LRO
in the whole system, which we may infer from transport measurements. For low $p$,
just above $T_{\rm c}$, the $z$-direction resistivity $\rho_c$ is $\approx 10^{3}-10^{6}$ larger 
than the $a$ or $b$-axis resistivity\cite{ResistBSLCO.2003,ResistC.ab.2004}.
Despite this large difference, it is surprising that both $\rho_c(T)$, and $\rho_{ab}(T)$ fall 
to zero at the same temperature ($T_{\rm c}$) and this puzzling behavior 
can be understood in terms of the Josephson coupling of Eq. \ref{EJ}. 
The smaller planar resistances yield larger $E_{\rm J}$ that promote first
LRO in the planes, but each plane $i$ would have its own SC phase $\theta_i$
if it was not for the weaker inter-plane $E_{\rm J}$ coupling.
Thus the $z$-direction Josephson coupling connects the planes, leading to only
a single-phase $\theta$ in the whole system, and both $z$ and $ab$ resistivity
drop off together at $T_{\rm c}$, which is a plausible explanation
for this long-known non-conventional result.
Again using Eq. \ref{EJ},
a smaller low temperature superfluid density is expected along the $z$-direction than along the plane,
which is confirmed by the measured large anisotropy of $ab$ and $c$-axis magnetic
penetration depth\cite{Penet.Depth.abc2000,YPenet.Depth.abc1999}. Thus, although $z$-direction coupling
is fundamental, the SC properties like the local SC amplitudes $\Delta_d({\bf r}_i, p,T)$
and $T_{\rm c}(p)$ develop and depend entirely on the CuO planes.

\section{the Superfluid Density}

For low-temperature superconductors,
the temperature $T_{\theta}$ at which LRO
disappears is very large compared with $T_{\rm c}$,
but they are estimated to be comparable for HTS\cite{Emery1995}. Our basic
point is that $T_{\rm c} = T_{\theta}$, and at higher temperatures,
the Cooper pairs in the nanoscopic grains have all different $\theta_i$ and,
consequently are in an incoherent state. This approach leads us to
infer a close connection between Josephson coupling and superfluid phase stiffness.

The first thing to
notice is that $J_{\rm c}(T)$ is proportional to the two-dimensional (2D) superfluid density
$n_{\rm s}(T)$ or the local Josephson current\cite{Spivak1991} 
that is also proportional to the phase stiffness 
$\rho_{\rm sc}$\cite{Bozovic2016}.
Along these lines, we made ${\left < E_{\rm J}(p,0) \right >}$ 
equal to the 2D zero-temperature superfluid phase stiffness\cite{Mello2020a,Mello2021} 
$\rho_{\rm sc}(p, 0)$ and reproduced the measured scale relation 
between the zero temperature superfluid density and $T_{\rm c}(p)$.
To underdoped compounds, this relation is known as Uemura's law\cite{Uemura1989},
and a similar relation for La-based overdoped films was obtained more
recently\cite{Bozovic2016}.

\begin{figure}[!ht]
 \includegraphics[height=6.0cm]{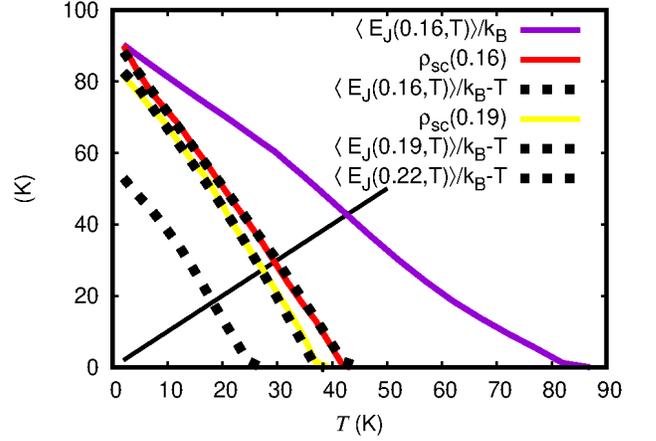}
\caption{ 
Comparison between the measured phase stiffness $\rho_{\rm sc} (p, T)$
for La-based films with $p = 0.16, 0.19$, and 0.22 from Ref.[\onlinecite{Bozovic2016}]) 
and $({\left < E_{\rm J}(p,T) \right >}/k_{\rm B}-T)$ (dashed lines), 
with an almost perfect agreement.
We show also the Josephson coupling ${\left < E_{\rm J}(0.16, T) \right >}$ that sets
the SC phase order scale $T_{\theta}$ derived in Ref. [\onlinecite{Mello2021}]
and the temperature $T$ to make clear how to derive 
$T_{\rm c}(p = 0.16)$.
}
\label{fig.1rho_s}
\end{figure}

Along these lines, we show in Fig. \ref{fig.1rho_s} that
$({\left < E_{\rm J}(p,T) \right >}/k_{\rm B}-T)$ reproduces the 
measured\cite{Bozovic2016} 2D phase stiffness $\rho_{\rm sc}(p, T)$ temperature dependence.
The plots show very good agreement with $p = 0.16$ and 0.19 films
according to Ref. [\onlinecite{Bozovic2016}]. For $p = 0.22$ we could
extract with certainty only the experimental values at $T = 0$ and $T_{\rm c}$, but
the theory seems to reproduce closely also the intermediate data.
In the plots we used ${\left < E_{\rm J}(p,T) \right >}$ from Eq. \ref{EJ}
with ${\left <\Delta_d(p,T)\right >}$ and $R_{\rm n}(p)$
derived previously in Refs.[\onlinecite{Mello2021},\onlinecite{Mello2020a}].

Another indication that ${\left < E_{\rm J}(p,0) \right >}/k_{\rm B} - T$ is 
equal to $\rho_{\rm sc} (p, T)$ comes from the proximity
effect experiment on YBCO wires\cite{ybco2000PE}. They measured the 
critical current $I_{\rm c}$ through a small insulator
region of size $d$ inserted in the SC wire. In their Fig. 2(b), they showed that
the product $I_{\rm c}R_{\rm n}$ at low-temperature is constant for 
several values of $d$ from 40 to 110 nm, which is much larger than the 
estimated coherence length $\xi_{\rm SC}$ of 9 nm\cite{ybco2000PE}. 
We interpret this result recalling that
the low-temperature Josephson current is\cite{Ketterson}
\begin{equation}
    I_{\rm J} = \frac{E_{\rm J}(0)}{\hbar /(2e)} = \frac{\pi \Delta_d(0)}{2e R_{\rm n}}
\label{IJ} 
\end{equation}

Following the above discussion, taking $I_{\rm J} = I_{\rm c} \propto {\left < E_{\rm J}(p,0) \right >}$
and according to Eq. \ref{EJ}, the product $I_{\rm c}R_{\rm n}$ is independent 
of the insulator spacer $d$, 
in perfect agreement with the measurements with 
the S-I-S YBCO junctions. This result, in the framework of
our theory, suggests that the CDW is also present in the 
underdoped insulator part of the YBCO wire, in agreement with the 
connection between CDW and the antinodal PG mentioned in the previous 
section\cite{Wise2009,Comin2014}.

\section{Results and Discussion}

In general, the critical current density is the product of the superfluid
velocity and density, that is, $J_{\rm c} = v_s \rho_{\rm sc}$.
We have shown above
that ${\left < E_{\rm J}(p,T) \right >}/k_{\rm B} - T$
is equal to $\rho_{\rm sc} (T)$ and that the low-temperature critical 
current $I_{\rm c}(0) \propto  {\left < E_{\rm J}(p,0) \right >}$ on
the YBCO wires\cite{ybco2000PE}. Based on
these two results, we assume that the average superfluid velocity is 
$v_s \propto {\left < E_{\rm J}(p,T) \right >}/{\left < E_{\rm J}(p,0) \right >}$
and that,
\begin{equation}
  J_{\rm c}(T) \propto \left( {\left< E_{\rm J}(p,T) \right >}/k_{\rm B} - T\right)
  \frac{{\left < E_{\rm J}(p,T) \right >}}{{\left < E_{\rm J}(p,0) \right >}} \;.
\label{IJc} 
\end{equation}
This expression  will be used
to compare with the GPE measured critical currents\cite{ybco2000PE,Bozovic2004}. 

The first system to deal here is the trilayers along the $z$ direction
composed of La$_{1.85}$Sr$_{0.15}$CuO$_4$ films with
$T_{\rm c} \approx 45$ K (S) and the underdoped 
superconductor La$_2$CuO$_{4+\delta}$ with $T'_{\rm c} \approx 25$ K (N') 
sandwiched according
to Ref.[\onlinecite{Bozovic2004}]. Initially we study
S and N' separately, with the average amplitudes 
${\left <\Delta_d(p= 0.12,T)\right >}$ and ${\left <\Delta_d(p= 0.15,T)\right >}$
\cite{Mello2020a,Mello2021} and the factor of 
three in their normal resistances $R_{\bf n}$ just above their $T_{\rm c}$ according to Ref. [\onlinecite{Bozovic2002}].  
For  $T \le 25$ K there is LRO in both S and N'.  
Therefore, in principle, $J_{\rm c}$ would be limited
by the lower $T'_{\rm c} \approx 25$ K, but they
measured a persisting $J_{\rm c}$ almost to $T = 35$ K independently of
the number of underdoped N' layers\cite{Bozovic2004}.

\begin{figure}[!ht]
 \includegraphics[height=6.0cm]{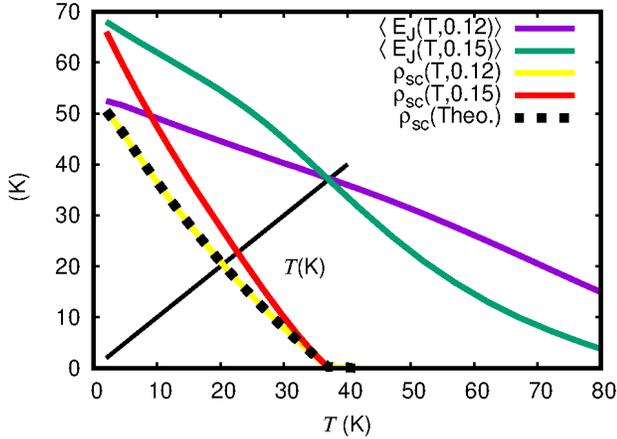}
\caption{ 
S and N' layers plots after they are connected in the S-N'-S structure
both with the average  hole density $p' = 0.135$.  
${\left < E_{\rm J}(p',T) \right >}$ for S (green) and N' (purple).
We also plot
$[{\left < E_{\rm J}(p',T) \right >}/k_{\rm B} - T] = \rho_{\rm sc} (p',T) $ for S (red)
and N' (yellow). The smaller $\rho_{\rm sc} (p',T)$ of N' is the maximum
critical current through the 
S-N'-S system [the superfluid density through S (red line) is larger] 
and is assumed to be proportional to the theoretical
critical current $J_{\rm c}$ (dashed line).
}
\label{Theoria2}
\end{figure}

The main reason to have supercurrents above $T'_{\rm c} = 25$ K is the  
presence of SC amplitudes $\Delta_d$ in the CDW 
charge domains without LRO,
what characterizes the PG region of the N' superconductor. 
Another important factor is the rearrangement of the holes between
the layers according to subsequent experiments: using similar 
La-based materials but combining undoped insulators with overdoped metallic 
La$_{2-x}$Sr$_{x}$CuO$_4$ with $x = 0.38$\cite{Smadici2009} and
$0.44$\cite{Suter2012}. The authors of those studies verified that the conducting holes pile up at the interface and redistribute themselves from the hole-rich to the insulator layers. Analysis with REXS\cite{Smadici2009}
and $\mu$-SR\cite{Suter2012} revealed that this reorganization of the holes 
occurs near the interface and they are not followed by the Sr dopant atoms.
The rearrangement of the hole densities add on or 
subtract carriers and
affect directly the local pair amplitudes $\Delta_d$, which depend on the local 
densities $p(i)$ as discussed above. 
The redistribution and uniformization of the charges will 
enhance ${ \left <\Delta_d \right >}$ in N' with a concomitant decrease in S.

\begin{figure}[!ht]
 \includegraphics[height=6.0cm]{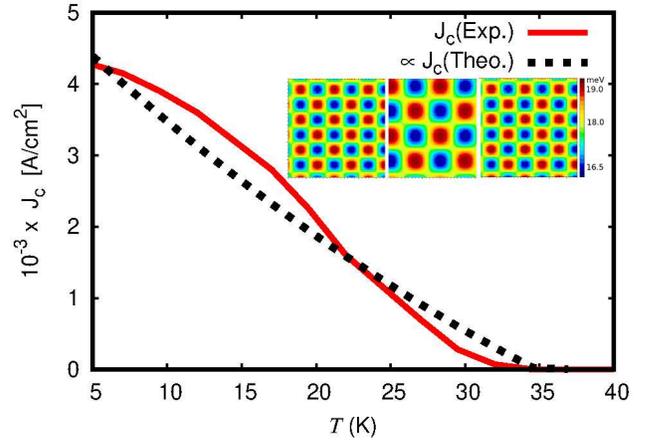}
\caption{ 
The solid red line is the S-N'-S experimental results 
from Fig. 2 of Ref. [\onlinecite{Bozovic2004}], and
the dashed line is our calculations that are proportional to $J_{\rm c}$.
We made both curves equal near $T = 5$ K because our estimates are more precise 
at low temperatures. The inset shows schematically
the coupled systems used in the calculations. In the trilayer 
experiments\cite{Bozovic2004,Morenzoni2011} the layers are on top of each other
and connected by a small Josephson coupling, but in the text, we explain that the
calculations are similar.
}
\label{TheoriaExp}
\end{figure}

This enhancement or weakening was observed in Meissner effects 
studies by low energy $\mu$-SR\cite{Morenzoni2011}. 
They showed that the local magnetic field ${\left < B_x\right >}$ 
along the $z$ direction in a similar
heterostructure S-N'-S is excluded like in a single
uniform superconductor even at temperatures more than 3 times larger than 
$T'_{\rm c}$. This reinforces our point that S and N' at $T'_{\rm c} < T < T_{\rm c}$
have both the same structure differing only by the presence or absence of LRO 
in the charge domains.

Now we will apply these ideas to the GPE in the La-based trilayer system\cite{Bozovic2016} .
The effect of the checkerboard charge inhomogeneities in the local SC properties were studied
in detail\cite{Mello2020a} in a system similar to N' with $p = 0.12$, and the simulations 
were shown above in Fig. \ref{MapsHist3}.
The $\Delta_d({\bf r}_i)$ amplitude histogram 
is shown in Fig. \ref{MapsHist3}(d), and the calculations show
that CDW higher (lower) local hole densities have larger (smaller) local SC amplitudes\cite{Mello2017}.
Therefore,  when a supercurrent flows,
the S-N'-S system acquires a mean hole density close to their average;
that is, the mean hole density in N' and S become  $p' \approx 0.135$.
This rearrangement increases 
${\left <\Delta_d (N', T)\right >}$ by a factor of approximately 1.5 and decreases 
${\left <\Delta_d (S, T)\right >}$ by a factor of approximately 0.6. Taking these changes into
account, we calculate the new Josephson couplings in N' and S
keeping their original $R_{\rm n}$ unchanged. The results of these calculations 
are shown in Fig. \ref{Theoria2}, where we plot these renormalized 
${\left < E_{\rm J}(p',T) \right >}$, that is with $p'= 0.135$ (green 
for S and purple for N') and  $\rho_{\rm sc}(p', T)$ 
for the N' (yellow) and S (red) coupled layers. Since the critical current
must be constant through the layers and N' holds the smaller values, we 
take exactly this (yellow) lower current as the theoretical critical current through the entire 
system (dashed line).

Since $J_{\rm c} = v_s \rho_{\rm sc}$, $v_s (p,T) \propto 
{\left < E_{\rm J}(p,T) \right >}/{\left < E_{\rm J}(p,0) \right >}$, and
using the dashed $\rho_{\rm sc}(p', T)$ curve on Fig. \ref{Theoria2}, 
we obtain our estimation to the
critical current measurements\cite{Bozovic2004} shown
in Fig. \ref{TheoriaExp}. To perform this comparison,
we made our results equal to the lowest measured temperature $T \approx 5$ K, 
where our estimation to $v_s ( T)$ is more precise.
Our results follow near
the experimental $J_{\rm c}(T)$ but do not have the round features displayed 
by the red curve that was taken from Ref. [\onlinecite{Bozovic2004}]. This is mainly
because of the poor estimation of $v_s(T)$ at finite temperatures.
Another reason is that the  S-N'-S system is formed with stacked layers along the 
$z$ direction and our calculations are with layers in the CuO plane 
as shown in the inset of Fig. \ref{TheoriaExp}. 
But as discussed before, the calculations follow along the same lines in the sense that LRO 
is attained first in CuO planes and, afterward,
by interlayer Josephson coupling, the whole system becomes superconducting.
Another possible correction is that 
the $z$-direction current may enhance influences of out-of-plane dopants or 
oxygen interstitials\cite{Campi2015}.

We argued that the formation of a granular superconductivity in the CO domains
is the most basic property of cuprate superconductors.
We showed that phase-ordering kinetics may describe electronic phase separation
transitions by the GL potential $V_{\rm GL}$ with
incommensurate free energy wells that originate the observed CDW. 
Charge fluctuations inside the domains may induce
hole-hole SC interaction proportional to  
the depth of the $V_{\rm GL}$ wells averaging over the whole system, that is, 
${\left < V_{\rm GL}(p)\right >}$\cite{Mello2017,Mello2021}.
The average
Josephson coupling ${\left < E_{\rm J}(p,T) \right >}$ between the local SC order parameter 
yields the LRO or phase-ordering temperature that is made equal to $T_{\rm c}$.
Furthermore, ${\left < E_{\rm J}(p,T) \right >} - k_{\rm B}T$ provides
perfect agreement
with the measured superfluid density temperature dependence of La-based overdoped 
films\cite{Bozovic2016}. The granular superconductivity is also 
the key mechanism behind the GPE because both S,  N', and I (with nonzero doping)
used in the experiments have all localized SC order parameters 
and differ only by the absence or presence of LRO. 

We should emphasize that the Josephson coupling
of Eq. \ref{EJ} is inversely proportional to the resistivity, which leads to much 
smaller $E_{\rm J}$ inter-plane coupling, and its identification with the local superfluid
density is confirmed by the anisotropic magnetic penetration length 
measurements\cite{YPenet.Depth.abc1999,Penet.Depth.abc2000}.
We finish pointing out that
the presence of CDW in electron-doped compounds\cite{Ndoped2018} is a further indication that
the method applies to all cuprates, and which we will explore further in the future.

\section{Acknowledgements}

We are grateful to David M\"ockli for critical reading the manuscript and 
acknowledge partial support by the Brazilian agencies CNPq and FAPERJ.

%

\end{document}